\documentclass[useAMS,usenatbib]{mn2e} 
\RequirePackage{lineno}
\usepackage{graphicx}  
\usepackage{amssymb}
\usepackage[usenames]{color}
\usepackage{array,multirow}

\usepackage[T1]{fontenc}
\usepackage{ae,aecompl}

\usepackage[a4paper,bottom=2cm,left=1.5cm,right=1.5cm,top=2cm]{geometry}
\usepackage{amsmath}	
\usepackage{subfigure}
\usepackage{listings}

\usepackage{xspace}
\usepackage{url}

\usepackage{times}
\usepackage[english]{babel}
\usepackage{mathrsfs}

\usepackage{threeparttable}
\usepackage{booktabs}

\usepackage{xcolor}
\definecolor{green}{HTML}{33CC33}
\definecolor{red}{HTML}{FF3300}
\definecolor{blue}{HTML}{3333FF}

\usepackage[format=plain,labelformat=default, labelsep=endash, labelfont=bf, margin=0pt, tableposition=top, font={footnotesize}]{caption}
\usepackage[colorlinks=true,citecolor=blue,urlcolor=green]{hyperref}

\usepackage{twoopt}
\usepackage[varg]{txfonts}
\usepackage[normalem]{ulem}

\renewcommand{\eqref}[1]{Equation~\ref{#1}}
\newcommand{\fref}[1]{Figure~\ref{#1}}

\newcommand{\sref}[1]{Section~\ref{#1}}

\newcommand{\aapr}{A\&ARv}
\newcommand{\aap}{A\&A} 
\newcommand{\apj}{ApJ} 
\newcommand{\apjl}{ApJL} 
\newcommand{\mnras}{MNRAS} 
\newcommand{\solphys}{Sol. Phys.} 
\newcommand{\nat}{Nature}

\newcommand{\aaps}{A\&AS} 

\newcommand{\ie}{i.e.\@\xspace} 
\newcommand{\eg}{e.g.\@\xspace} 

\newcommand{\numax}{\ensuremath{\nu_{\rm max}}\xspace}

\numberwithin{equation}{section}

\makeatletter
\def\maketag@@@#1{\hbox{\m@th\normalfont\normalsize#1}}
\makeatother

\makeatletter
\DeclareRobustCommand*\textsubscript[1]{%
  \@textsubscript{\selectfont#1}}
\def\@textsubscript#1{%
  {\m@th\ensuremath{_{\mbox{\fontsize\sf@size\z@#1}}}}}
\makeatother

\makeatletter
\newcommand*\mysize{%
  \@setfontsize\mysize{5.7}{8.0}%
}
\makeatother

\makeatletter
\newcommand*\tabsize{%
  \@setfontsize\tabsize{7.}{8.0}%
}
\makeatother

\usepackage{blindtext}
\usepackage{lipsum} 

\title[2B coherence analysis]{Spatial incoherence of solar
  granulation: a global analysis using BiSON 2B data}

\author[M. N. Lund et al.]{Mikkel~N.~Lund$^{1,2}$\thanks{E-mail: \href{mailto:mikkelnl@phys.au.dk}{mikkelnl@phys.au.dk} (MNL)},
William~J.~Chaplin$^{1,2}$,
Steven~J.~Hale$^{1,2}$,
Guy~R.~Davies$^{1,2}$,\newauthor
Yvonne~P.~Elsworth$^{1,2}$,
Rachel~Howe$^{1,2}$
\vspace*{0.5em} \\ 
$^1$School of Physics and Astronomy, University of Birmingham, Edgbaston, Birmingham, B15 2TT, UK\\
$^2$Stellar Astrophysics Centre, Department of Physics and Astronomy, Aarhus University, Ny Munkegade 120, DK-8000 Aarhus C, Denmark\\
}

\begin{document}

\vspace{-10cm}
\date{Accepted 2017 August 21; Received 2017 August 21; in original form 2017 February 19}
\pagerange{\pageref{firstpage}--\pageref{lastpage}} \pubyear{2016}

\maketitle

\label{firstpage}
\vspace{-3cm}
\begin{abstract}
A poor understanding of the impact of convective turbulence in the outer layers of the Sun and Sun-like stars challenges the advance towards an improved understanding of their internal structure and dynamics. Assessing and calibrating these effects is therefore of great importance.
Here we study the spatial coherence of granulation noise and oscillation modes in the Sun, with the aim of exploiting any incoherence to
beat-down observed granulation noise, hence improving the detection of low-frequency p-modes. Using data from the BiSON 2B instrument, we assess the coherence between different atmospheric heights and between different surface regions. We find that granulation noise from the different atmospheric heights probed is largely incoherent; frequency regions dominated by oscillations are almost fully coherent. We find a randomised phase difference for the granulation noise, and a near zero difference for the evanescent oscillations. A reduction of the incoherent granulation noise is shown by application of the cross-spectrum.
\end{abstract}


\section{Introduction}

Helioseismology and asteroseismology are playing a fundamental role in
the quest for an improved understanding of the internal structure and
dynamics of the Sun and Sun-like stars. 
A lack of understanding of the influence that convective
turbulence in the outer layers of the stars has on the oscillation
frequencies does, however, hamper some aspects of these studies.

The near-surface turbulence in stars is very difficult and
time-consuming to model, and it systematically shifts the frequencies of
the oscillations \citep[see,
  \eg,][]{1999A&A...351..689R,2013EAS....63..367M,2014MNRAS.437..164P,2017MNRAS.464L.124H,2017MNRAS.466L..43T}. From
solar data we know that the shifts are stronger in high-frequency
modes, which with their higher upper turning point are more sensitive to near-surface effects, compared to modes of
lower frequency \citep[][]{2008ApJ...683L.175K,2016A&A...592A.159B}.
One approach to addressing the problem is the detection and use of
very-low frequency modes, including core-sensitive g-modes, as these
are much less affected by the near-surface turbulence. Such modes may
then be used as ``anchors'' to properly calibrate the effects, since
they in principle provide diagnostics largely free from the impact of
the near-surface layers. The low-frequency modes are also important
because they, from their higher frequency precision and sensitivity towards deeper stellar layers, can provide exquisite information on the overall internal and differential rotational
properties of stars \citep[][]{2014ApJ...790..121L}. This information is vital for improving our
understanding of the dynamical processes driven by rotation, such
as solar and stellar activity cycles
\citep[][]{2007MNRAS.377...17C,2010Sci...329.1032G}, and the transport
of energy, angular momentum, and chemical species in stars ---
ingredients that currently are poorly described in stellar evolution
models.

The low-frequency modes do, however, show very low amplitudes, and must
be detected against a background noise whose dominant contribution is
intrinsic solar or stellar noise due to convection (specifically the signature of granulation, which is the surface
manifestation of convection). The stochastic nature
of the convection implies that any signals from it will be spatially
incoherent over some characteristic lengthscale, both with depth in
the stellar atmosphere and with position on the surface. Our study
here is motivated by the aim of exploiting this incoherence to
beat-down the observed granulation noise, and to thereby improve
observed signal-to-noise ratios in low-frequency p-modes
\citep[][]{2009MNRAS.396L.100B,2014MNRAS.439.2025D}, and solar g-modes
\citep[][]{2007Sci...316.1591G,2010A&ARv..18..197A,2010MNRAS.406..767B}.
The oscillation modes of interest are expected to be largely coherent
because they are standing modes of the acoustic cavity formed by the star. We look in detail at
the spectral coherence and phase of disk-averaged Doppler velocity
observations of the Sun that have different sensitivies in atmospheric
height and spatial location on the visible disk. These data were
collected by the ``2B'' instrument of the Birmingham Solar
Oscillations Network (BiSON)
\citep[][]{1996SoPh..168....1C,2005MNRAS.359..607C,2016SoPh..291....1H}. 

The ability to combine observations from different depths to reduce
granulation noise is in fact a foundational idea in the development of
the successor to the Global Oscillations at Low Frequency
\citep[GOLF;][]{1995SoPh..162...61G} instrument, GOLF-NG \citep[New Generation; see][]{2006AdSpR..38.1812T,2012ASPC..462..240T}. Our study can
be seen as a complement to the analysis of data from GOLF by
\citet[][]{2004ESASP.559..432G}, which suggests that the idea of
incoherent granulation noise is sound. Our observations are of much
longer duration, and use a different solar spectral line.

The study of different atmospheric heights can also be used to
determine the depth dependence of properties like granulation and
oscillation mode amplitudes, and mode phases. Crucially, the
depth-dependent properties of granulation could be used to calibrate
detailed 3D convection models
\citep[][]{1995A&AS..109...79E,2008A&A...490.1143L}. In a coming
publication we will study the depth-dependence of various properties
from the 2B data, but in the current work we focus on the coherence
and phase at different depths for a reduction of incoherent signals.

The paper proceeds as follows: In \sref{sec:inst} we describe the
basic observing strategy of BiSON and the 2B instrument;
\sref{sec:data} presents the data collected. In \sref{sec:sc} we
outline the method of spectral coherence and phase, and present results for the 2B data in \sref{sec:res}. In \sref{sec:dis} we
discuss our findings and their impact on future proposed solar
observations.

\section{The 2B instrument}
\label{sec:inst}

\begin{figure}
\includegraphics[width=\columnwidth]{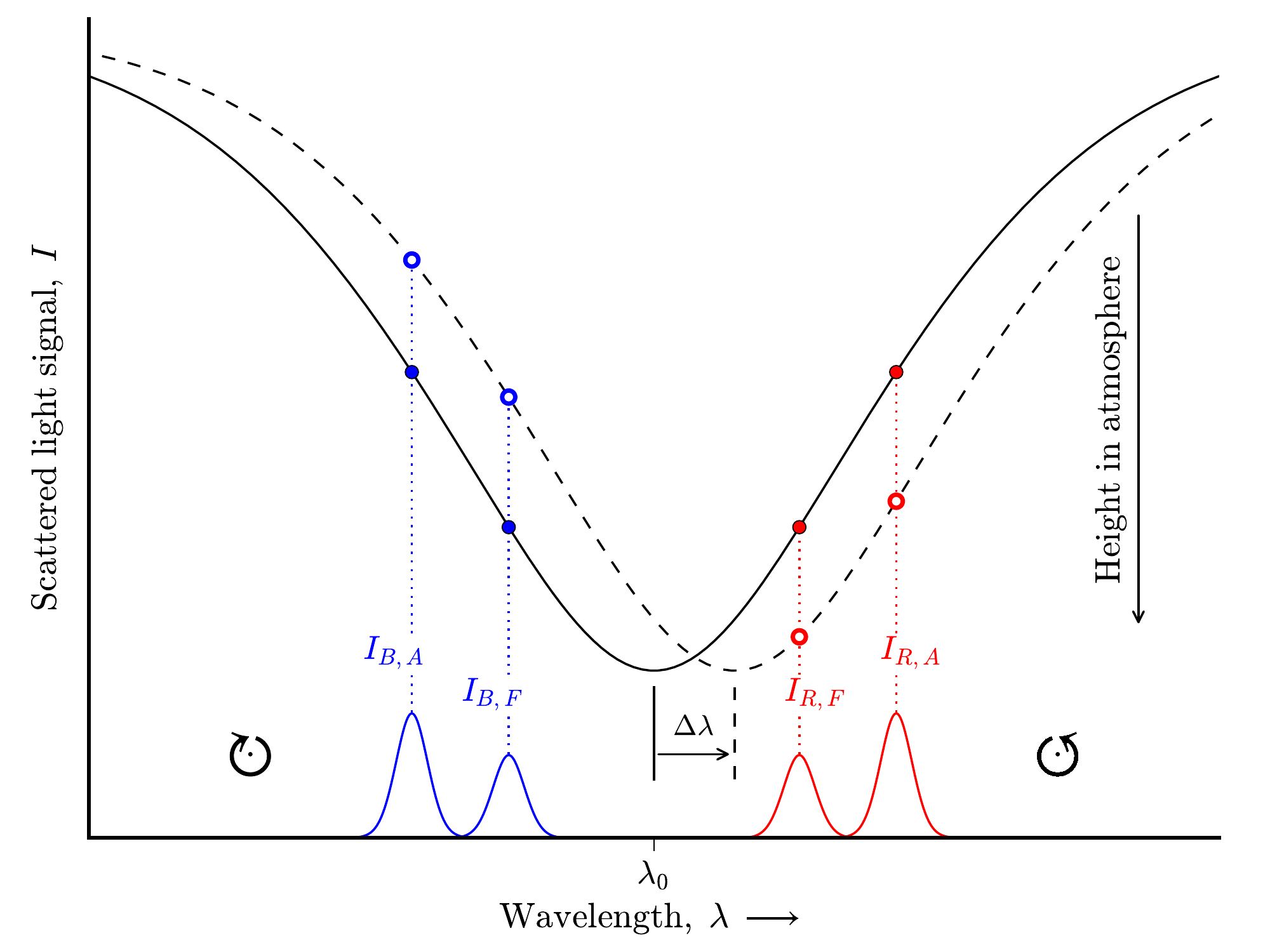}
\caption{Schematic of the D1-line of K\textsc{i}, showing the principle of the BiSON measurements. The full black line shows the rest frame K\textsc{i} profile centred on $\rm\lambda_0$, while the dashed line shows a displaced version (shifted by $\rm\Delta\lambda$) due to a positive velocity. The working measurement points are indicated on the profiles, with the subscripts indicating if the point measured the intensity on the blue ($B$) or red ($R$) wing of the line, and whether the measurement is made by the forward ($F$) or aft ($A$) cell. Below each working point are shown schematic representations of the instrumental scattering responses. Shown on either side of the line center is also the handedness of the polarized light reaching the two allowed D1 transitions. The closer to the continuum level the line is measured, the deeper in the photosphere (smaller $H$) the velocity is probed. A measurement made towards the line centre measures higher up in the astmophere, \ie, further away from the base of the photosphere.  }
\label{fig:line}
\end{figure}
\begin{figure*}
\includegraphics[height=0.9\textwidth, angle=90]{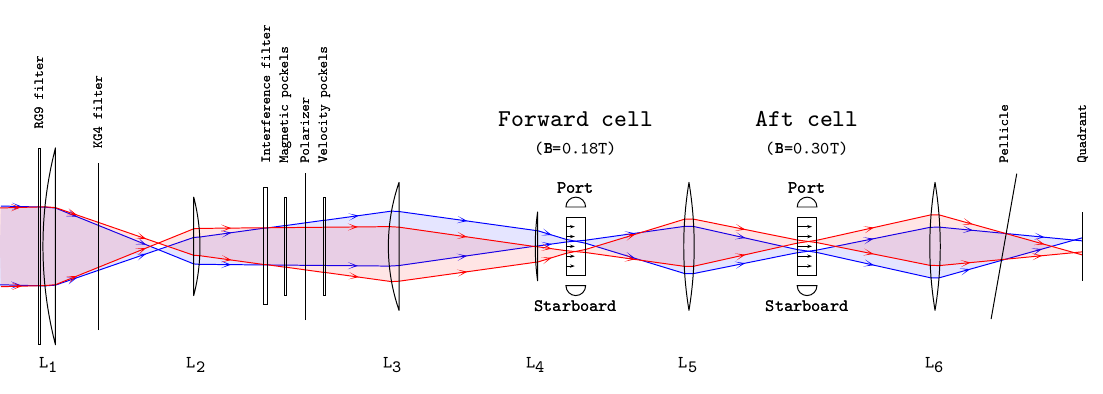}
\caption{Schematic of the optical setup for the 2B instrument. Light enters from the left, where the red (blue) beam indicates light from the receding (approaching) limb of the Sun. The different lenses have names from \texttt{L\textsubscript{1}}-\texttt{L\textsubscript{6}} given at the bottom of the schematic; other components, such as filters and pockels cells, have been indicated with their name. Shown are also the two potassium vapour cells (forward and aft), each sitting in a longitudinal magnetic field of different strength (indicated by the horizontal arrows with the cells). The two detectors for each cell (port and starboard) are also depicted.  }
\label{fig:setup}
\end{figure*}
BiSON employs as the standard instrument a single resonant scattering spectrometer (RSS), which measures Doppler shifts of the D1 Fraunhofer line of neutral potassium (K\textsc{i}), at 769.9 nm \citep[][]{1961Natur.189..373I,1976Natur.259...92B}.
By placing the potassium vapour cell in a longitudinal magnetic field, with a strength of around $0.18\rm T$, the energy levels of the $\rm 4^2S_{\frac{1}{2}}$ and $\rm 4^2P_{\frac{1}{2}}$ states are Zeeman split. 
The energy shifts result in a wavelength splitting of the two allowed D1 transitions of K\textsc{i}. In effect, the intensity may be measured in the red ($I_R$; longer wavelength) and blue ($I_B$; shorter wavelength) wings of the potassium line (see \fref{fig:line}). 
For a positive velocity shift as shown in \fref{fig:line}, a working point on the blue wing then probes the line profile closer to its continuum level, which is deeper (closer to the base of the photosphere) in the solar atmosphere; the red wing working point probes the part of the profile closer to the centre of the line, \ie, higher up in the solar atmosphere.
The two allowed transitions are sensitive only to circularly polarised light, and each only to a given handedness. By switching between the handedness of the incoming light using a polarising filter in front of the vapor cell, at a frequency of $100$Hz, the intensities of the two wings may be measured near simultaneously.
From the red and blue wing intensities the ratio
\begin{equation}
\mathcal{R}= \frac{I_B - I_R}{I_B + I_R}\, ,
\end{equation}
is used as a proxy for the Doppler shift, and hence the radial velocity at the height of formation in that part of the K\textsc{i} line.
The calibration of a velocity from $\mathcal{R}$ follows the procedure by \citet[][]{1995A&AS..113..379E}. Here, the correspondence between the measured $\mathcal{R}$ and the known velocity components from the Earth's orbit and spin are used to map $\mathcal{R}$ to a velocity. Corrections for the effects of differential extinction at the beginning and end of a given day are applied following \citet[][]{2014MNRAS.441.3009D}.

The 2B instrument is a double-field scattering spectrometer (see \fref{fig:setup}), that is, it employs two potassium cells placed in different magnetic field strengths \citep[][]{2BLewis}. 
In many aspects this resembles the setup of the GOLF instrument, which was built to measure the intensity 
of the neutral sodium doublet at four working points by resonant scattering in a single vapour cell, which 
then use a modulated magnetic field \citep[][]{1994A&AS..107..177B,1995SoPh..162...61G}.
2B is installed at the BiSON node in Carnarvon, Western Australia, and has been collecting data since its commissioning in 1991 as an extension to the basic RSS (known as Mark-V).
The first, or \emph{forward}, cell along the optical path is placed in a field of $0.18\rm T$ and is responsible for producing the standard output from the Carnarvon node.
The second, or \emph{aft}, cell is placed in a field of $0.3\rm T$.

2B was designed with two cells placed in different fields to enable analysis of solar mode and granulation properties at different heights in the solar atmosphere. The use of the same instrument to probe different heights has the advantage that most instrumental features will be shared, hence allowing a better isolation of the height dependence of physical phenomena in the Sun.
The sensitivity of the $\mathcal{R}$-ratio to velocity changes is a function of both velocity and magnetic field strength. During the year the minimum and maximum line-of-sight velocity expected $(v_{\rm orb} + v_{\rm spin} + v_{\rm grs})$ falls in the range from $-328\rm\, m/s$ to $1592\rm\, m/s$. Here $v_{\rm orb}$ and $v_{\rm spin}$ denote the velocity components from the Earth's orbit and rotation, and $v_{\rm grs}$ gives the contribution from gravitational redshift.
Concerning the specific field strengths adopted, the field of $0.18\rm T$ is chosen as the 
standard for the basic BiSON RSS, because it samples the solar K\textsc{i} line where its slope 
is largest, thereby maximising the sensitivity of $\mathcal{R}$ to velocity shifts of the line in the expected velocity range from observations, and this high sensitivity is maintained through the range of $\pm 2\rm \, km/s$. 
The $0.3\rm T$ field of the aft cell results in a slightly reduced sensitivity and was chosen based on a compromise between several competing qualities --- 
a large height differential $\Delta H$ is desired compared to the working points of the 
$0.18\rm T$ field, which suggests sampling the K\textsc{i} line further out in the wings, 
corresponding to applying a higher magnetic field; a high sensitivity of $\mathcal{R}$ to line shifts 
is desired for good quality measurements, which suggests moving closer to the $0.18\rm T$ fields where 
the sensitivity is maximised; large values for the amplitudes of the solar oscillations are desired for 
detection of low S/N modes and in general for precise mode analysis, which suggests moving higher in the 
atmosphere and thus closer to the line centre, corresponding to a lower magnetic field strength. 
A last consideration is that the overlap between the Zeeman split lines should be minimised. Only then can the measurements by the forward and the
aft cells be considered independent. Using $B=0.3\rm T$ for the aft cell gives an overlap of only ${\sim}0.2\%$. The difference in height of formation
of the part of the D1-line probed by the aft and forward cells is then approximately $\rm {\sim}165\,km$ \citep[][]{2BLewis}, based on a simple calculation of
the height where the monochromatic optical depth is unity \citep[][]{1996SoPh..163..231U}.
\fref{fig:line} shows the principle of the intensity measurements of the 2B instrument.

Each cell has two detectors, referred to as the \emph{port} and \emph{starboard} detectors, sitting on either side of the cells.
The image of the Sun has a finite extent of about $4\rm mm$ in the centre of the vapor cells, which leads to an effect referred to as instrumental Doppler imaging --- a given detector will have a higher sensitivity to the signal from the closest solar limb \citep[][]{2009MNRAS.397..793B}, because the intensity of light from a given region of the Sun will have to travel a given path length through the vapor subject to extinction. As a result, the two detectors will have different spatial sensitivities. 
The Doppler imaging also has an important contribution from the sensitivity weighting from the solar differential rotation and overall limb darkening \citep[see][]{1978MNRAS.185...19B,GuyThesis,2014MNRAS.441.3009D} --- the blue wing of the spectral line is heavily weighted towards the approaching limb of the solar disk; the red wing to the receding limb. The spatial weighting will further depend on the varying size of the solar image and the tilt and position angle of the solar rotation axis through the year.

We also note that the solar image is inverted between the two cells (see \fref{fig:setup}). So, where the highest sensitivity to, \eg, the approaching limb of the solar disk is obtained by the port detector in the forward cell, it will in the aft cell be obtained by the starboard detector. The port (or starboard) detectors of the two cells will thus be spatially sensitive to different parts of the sun and to different atmospheric heights.

\section{Data}
\label{sec:data}

\begin{figure*}
\includegraphics[width=\textwidth]{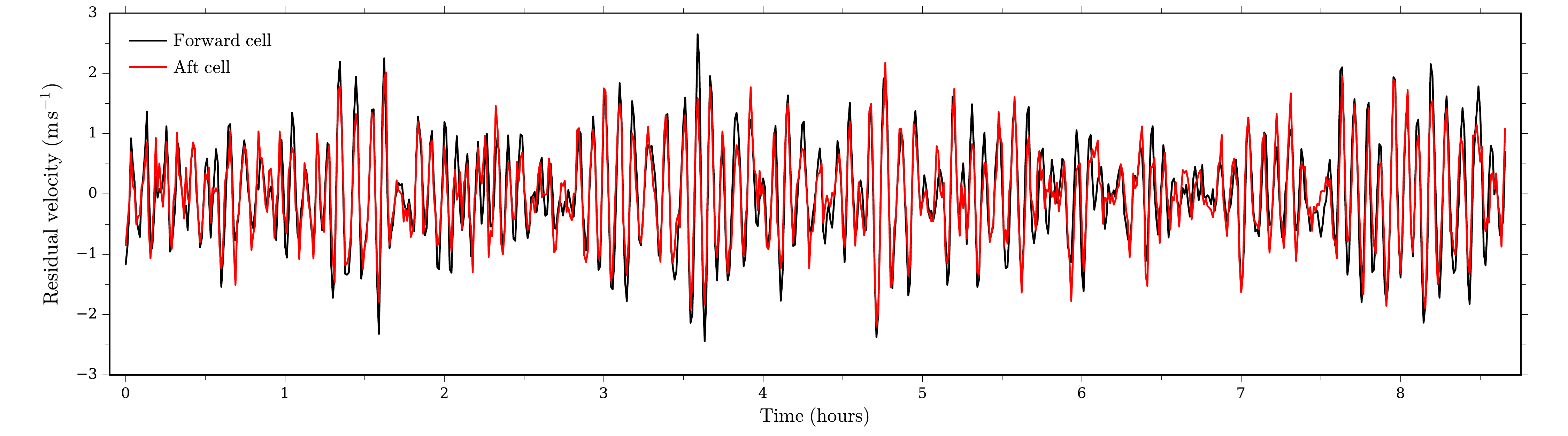}
\caption{Residual velocity time series for the forward and aft spectrometer cells from 8 June 2011. The time series from each cell is constructed from the mean of the port and starboard detectors of the cell.}
\label{fig:ts}
\end{figure*}

\begin{figure}
\includegraphics[width=\columnwidth]{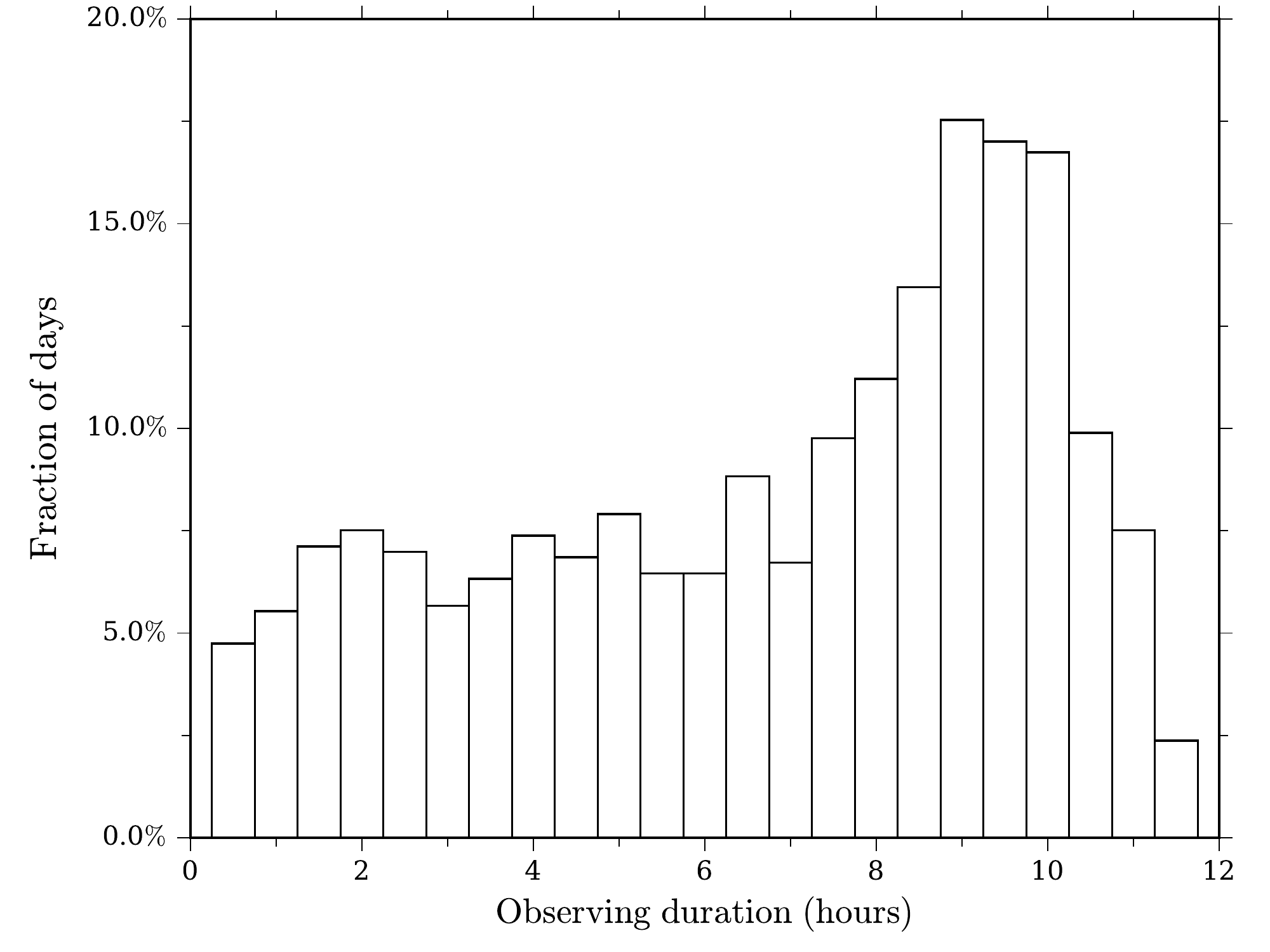}
\caption{Observing duration per day for data collected between 11 July 2009 and 21 April 2016, counting only days where some observations were obtained.}
\label{fig:fill}
\end{figure}

In this study we use data collected between 11 July 2009 and 21 April 2016, with a sampling rate of 40 seconds. \fref{fig:ts} gives an example of the residual velocity data during a given day for the forward and aft cells. For each cell we obtain the residual velocity signals separately from the \emph{port} and \emph{starboard} detectors, together with their average. The data from the commissioning of the instrument until the start date used here exhibit many instrumental artifacts and interruptions in the data collection due to technical issues. 
For the adopted period, spanning $1826$ days, data were collected during $1524$ days with a ${\sim}28\%$ 
duty cycle (${\sim}429$ days), with a modal value of $9-10$ hours of data per day (see \fref{fig:fill}). 
As we are looking at data from a single station the data gaps are dominated by a diurnal cycle, and 
secondly by weather and instrumental issues. 
For the above statistics we restrict ourselves to the times where both cells are recording  (\eg, excluding times when one cell is down for technical reasons).
In addition, we only use daily segments with more than 5 hours of data. We note that the velocity calibration from $\mathcal{R}$ \citep[][]{1995A&AS..113..379E,2014MNRAS.441.3009D} will attenuate low frequency signals; with a minimum daily data length of 5 hours it is expected that signals $\rm {\lesssim}200\, \mu Hz$ will be affected to some degree \citep[][]{GuyThesis}. 

\section{Spectral and phase coherence}
\label{sec:sc}

The coherence is useful for describing the relationship between two data sets in the frequency domain, and represents the part of the signal in a time series A that can be explained by the signal in a time series B at a given frequency. For spectral coherence we will look at the magnitude squared coherence (MSC), which then gives the fractional power at a given frequency that can be explained by time series A's linear regression on B. The MSC gives a value in the interval $0\leqslant {\rm MSC} \leqslant 1$, where $0$ indicates no coherence and $1$ full coherence. Note, however, that given the hard boundaries of 0 and 1 for the MSC any noise will always offset the measured MSC value away from 0 and 1, even if the signals are fully coherent/incoherent. 

One may from the cross-spectrum analysis look at the phase difference between the two signals at a given frequency.
Note that if a fixed phase difference exists between the two signals they can still have a high coherence. This is different from time averaging where a phase difference between two signals will lead to cancellation even if the signals are coherent.

We construct the MSC from the coefficients of a least-squares sine-wave fit \citep[][]{1992PhDT.......208K,1995A&A...301..123F}, normalised by the root-mean-square (RMS) scaled version of Parseval's theorem. 
The following short-hand notations are used:
\begin{align}\begin{split}\label{eqn:summer}
	s_j &\equiv \sum\nolimits_i w_i \cdot x_i \cdot \sin(2\pi\nu_j t_i) \\
	c_j &\equiv \sum\nolimits_i w_i \cdot x_i \cdot \cos(2\pi\nu_j t_i) \\
	ss_j &\equiv \sum\nolimits_i w_i \cdot \sin^2(2\pi\nu_j t_i) \\
	cc_j &\equiv \sum\nolimits_i w_i \cdot \cos^2(2\pi\nu_jt_i) \\
	sc_j &\equiv \sum\nolimits_i w_i \cdot \sin(2\pi\nu_j t_i) \cdot \cos(2\pi\nu_j t_i)\, ,
\end{split}
\end{align}
where $x_i$ refers to the time series signal at time $t_i$; $w_i$ refers to the weight applied to the specific data point, and may be used to window the function (cf. Welch's method, see below).
We then define:
\begin{align}\label{eq:alphabeta}
	\alpha_j &= \frac{s_j \cdot cc_j - c_j \cdot sc_j}{ss_j \cdot cc_j - sc_j^2} \, , \\
	\beta_j &= \frac{c_j \cdot ss_j - s_j \cdot sc_j}{ss_j \cdot cc_j - sc_j^2} \, ,
\end{align}

Replacing ``$x$'' in \eqref{eqn:summer} with the forward (F) and aft (A) cell time series gives the auto, or power, spectra for these as: 
\begin{align}
P_{FF}(\nu_j) &= \alpha_{jF}^2 + \beta_{jF}^2 \\
P_{AA}(\nu_j) &= \alpha_{jA}^2 + \beta_{jA}^2 \, .
\end{align}

To define the cross-spectrum functions we construct the so-called co-spectrum:
\begin{equation}
C(\nu_j) = \alpha_{jF}\alpha_{jA} + \beta_{jF}\beta_{jA}\, ,
\end{equation}
and quadrature-spectrum:
\begin{equation}
Q(\nu_j) = \alpha_{jF}\beta_{jA} - \alpha_{jA}\beta_{jF}\, .
\end{equation}

The cross-amplitude spectrum at frequency $\nu_j$ between the forward and aft cell time series is then given by:
\begin{equation}
\langle P_{FA}(\nu_j) \rangle = \sqrt{\hat{C}^2(\nu_j) + \hat{Q}^2(\nu_j)}\, .
\end{equation}
The angle brackets, and hats on $Q$ and $C$, indicate that it is the expectation values for the respective spectra that are used in calculating the cross-spectrum.
Estimation of the expectation values for the spectra is typically done using Welch's method \citep[][]{1161901}, where the time series is divided into a given number of segments with a given overlap. Each segment is then windowed using a taper function (often a Hanning window) to minimise edge effects, and the cross- and auto-spectra are then calculated for each segment. The expectation values are then obtained as the average of the spectra from all the segments. 
One might also approximate the expectation value spectra by applying smoothing to the cross- and auto-spectra across a given number of frequency bins \citep[][]{Pardo2012207}.

The magnitude squared coherence (MSC) is then given as \citep[][]{chatfield2004timeseries}:
\begin{equation}\label{eq:msc}
{\rm MSC}(\nu) = \frac{|\langle P_{FA}(\nu)  \rangle|^2}{\langle P_{FF}(\nu)  \rangle\langle P_{AA}(\nu)  \rangle}\, .
\end{equation}
Note that if only a single realisation of the two time series are used, the MSC will always return a value of $1$, regardless of the actual coherence between the time series --- only by using the expectation values for the components does the MSC return a sensible value for the coherence.
The phase-spectrum, giving the phase difference between the two signals at a given frequency, can be computed as:
\begin{equation}\label{eq:phase}
\Delta\phi_{FA}(\nu_j) = \tan^{-1}\left(-\hat{Q}^2(\nu_j) / \hat{C}^2(\nu_j) \right)\, .
\end{equation}

The use of the cross spectrum for the reduction of incoherent signals is not new in solar observations.
To date the coherence has primarily been studied in the context of reducing incoherent instrumental noise 
and noise from the Earth's atmosphere. \citet[][]{1994MNRAS.269..529E} showed, \eg, a reduction in 
noise from joining data from two different BiSON stations at almost the same 
longitude (Iza\~{n}a and Sutherland), thus probing the same height in the solar atmosphere, 
but reducing incoherent contributions from instrumental and atmospheric noise. \citet[][]{1998ApJ...504L..51G} used 
the cross spectrum to reduce (high-frequency) photon noise from the GOLF photomultiplier tubes, 
enabling the detection of solar high-frequency pseudo-modes above the solar cutoff frequency.
More recently \citet[][]{2014MNRAS.441.3009D} showed a noise reduction from using contemporaneous 
data from BiSON stations at different longitudes. The difference in line-of-sight velocities of the 
stations (from the Earth's rotation) results in a probing of different atmospheric heights. 
\citet[][]{2014MNRAS.441.3009D} found less than full coherence in spectral regions not dominated by 
oscillations. This is similar to what we test here using 2B data (see \sref{sec:res}), 
but their results are unavoidably more ambiguous because the durations of contemporaneous 
observations between two (or three) specific stations are generally very short. 
See also \citet[][]{2007MNRAS.379....2B} for the use of contemporaneous data from two different 
instruments by formulating a joint probability in searching the combined data sets for excess power from 
oscillation modes.

\section{Results}
\label{sec:res}

With 2B we have the advantage that data are obtained for different depths at every sampling. In addition, because of the different spatial sensitivity of the port and starboard detectors of each cell, these preferentially observe different parts of the Solar surface. This effectively reduces the coherence of observed signals at a given depth, \ie, between the detectors of a given cell.
Below we present the results obtained for the analysis of the spectral coherence (\sref{sec:res_spec}) and phase difference  (\sref{sec:res_pha}) at different depths, and the resulting reduction of incoherence noise via the cross spectrum (\sref{sec:res_cro}). 

\subsection{Spectral coherence}
\label{sec:res_spec}
\begin{figure*}
    \centering
    \begin{subfigure}
        \centering
        \includegraphics[width=0.49\textwidth]{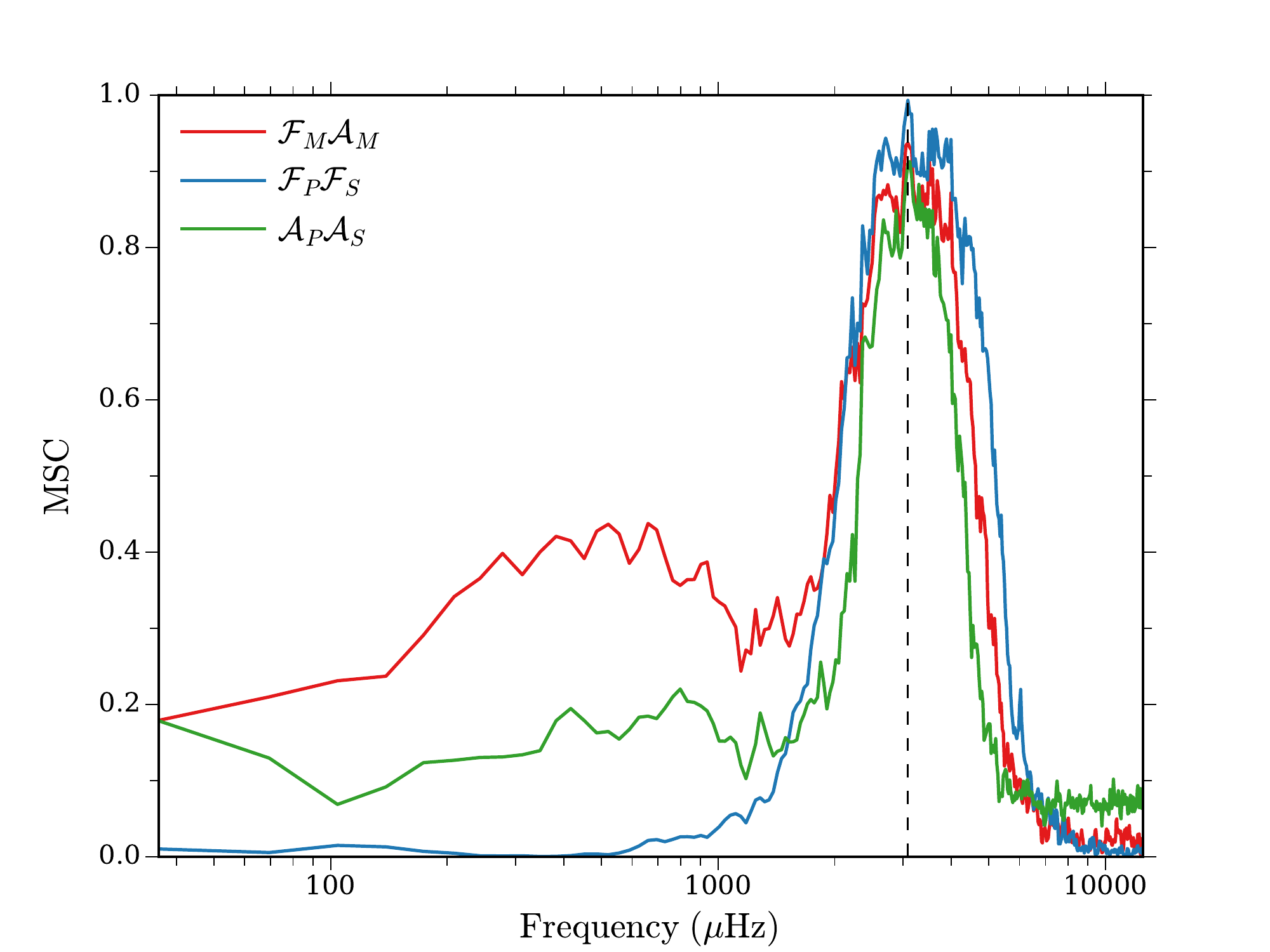}
    \end{subfigure}
    \hfill
    \begin{subfigure}
        \centering
        \includegraphics[width=0.49\textwidth]{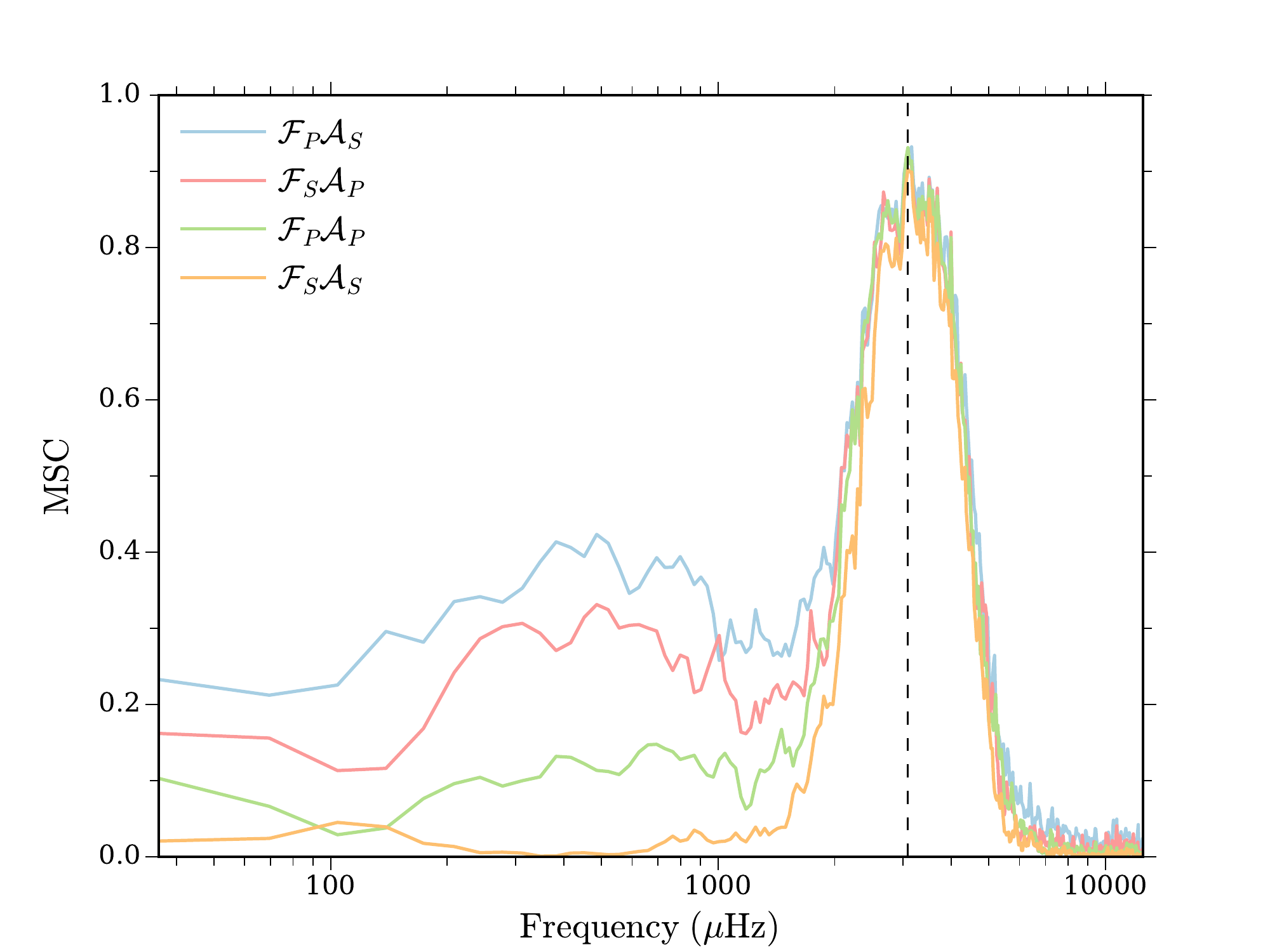}
    \end{subfigure}
    \caption{Magnitude squared coherence (MSC) functions from different 2B time series combinations as a function of frequency. The vertical dashed lines indicate the solar \numax of $\rm 3150\, \mu Hz$. Left: MSC functions from the forward ($\mathcal{F}$) and aft ($\mathcal{A}$) cells, combining the signals from the port ($P$), starboard ($S$), or mean ($M$) detector data products (see legend). Right: MSC functions from cross-combinations between the detectors of the two cells.}
\label{fig:msc}
\end{figure*}

In our study of the spectral coherence we are primarily interested in reducing the bias and scatter, at the cost of a reduced frequency resolution.
A natural segmentation of the time series thus comes from the diurnal nature of the observations. The use of daily segments in the calculation of expectation values alleviates the impact from the daily spectral window. We calculate the average spectra from the daily segments at the same resolution and without overlaps \citep[referred to as Bartlett's method;][]{BARTLETT01061950}, and here only include days where observations were conducted for more than five hours.  To avoid edge effects each daily segment was tapered using a Dolph-Chebyshev window \citep[][]{1697083,1455106} --- this effectively reduces the resolution, but has the benefit of also reducing the variance of the periodograms.

\fref{fig:msc} shows the MSC functions from the combination of different 2B time series --- the legend indicates whether the time series originate from the forward ($\mathcal{F}$) or aft ($\mathcal{A}$) cell; the subscript denotes the detector used in the given cell, \ie, using either the port ($P$), starboard ($S$), or mean ($M$) detector data products.

Overall, the different time series combinations, with contributions from different depths and/or spatial regions, show the same primary features: at frequencies below ${\sim}1500\, \rm \mu Hz$, the region dominated by the granulation background, the signals are largely incoherent ($\rm MSC\lesssim0.4$); at frequencies dominated by the stellar oscillations, ${\sim}1500$ to ${\sim}4500\, \rm \mu Hz$, they are, as expected, largely coherent ($\rm 0.4\lesssim MSC<1$); at high frequencies, ${\gtrsim}4500\, \rm \mu Hz$, where the signals are dominated by white noise the signals are incoherent ($\rm MSC\lesssim0.1$), although there will be some signal present due to pseudo-modes. As expected we see that signals $\rm {\lesssim}200\, \mu Hz$ are attenuated by the velocity calibration \citep[][]{GuyThesis}

Concerning the individual combinations, the left panel of \fref{fig:msc} shows the MSC functions from the same cells but different detectors, which isolates the effect from the surface weighting.
We do see a difference in these surface sensitive MSC functions for the forward and aft cells, \ie, at different depths. In other words, the different parts of the surface appear to have a lower coherence at the atmospheric height probed by the forward cell compared to the aft. 
We suspect that most of the difference seen between the two cells is of instrumental origin, and that the coherence level for the detectors of the aft cell is elevated across all frequencies compared to the forward cell. This elevated level is most clearly seen towards high frequencies for the aft cell --- one explanation for the origin of the signal could be non-resonantly scattered light in the optical system after the forward cell (see \fref{fig:setup}), hence adding a white correlated signal component to both detectors of the aft cell.
Another possible contribution arises from a slightly different overlap in the spatial weighting between the detectors at different depths --- this is caused by a difference in the correlation between velocity and intensity signals at different depths; in the overshoot region highest in the atmosphere, which is probed predominantly by the forward cell, the vertical velocity and temperature-contrast has the opposite correlation compared to that in the convection zone, which is closer to the region probed by the aft cell.

The right-hand panel of \fref{fig:msc} shows the MSC functions from the different cross-combinations between the detectors of the two cells. The most interesting feature seen here is the difference between using the detectors on a given side (\ie, port or starboard) for both cells or detectors on different sides. The combinations with the detectors on the same side are seen to have a lower coherence than the ones that mix the detector sides. The reason for this is the reversal of the solar image between the two cells (cf. \sref{sec:inst}), so in addition to the different depths probed by the two cells, the combinations adopting the detectors on the same side will include the reduction in coherence from weighting to different spatial regions of the solar surface. In both the MSC functions in the right panel, and the $\mathcal{F}_M\mathcal{A}_M$ combination, we see that the coherence shows a minimum around $\rm {\sim}1500\, \mu Hz$, but it increases slightly towards lower frequencies, until the attenuation from the velocity calibration sets in. This frequency dependence could be explained by a combination of common instrumental signals, noise from active regions, and the larger scales of convective structures (\eg, meso- and supergranulation) probed towards lower frequency \citep[][]{1985ESASP.235..199H,1994SoPh..152..247A}.

\subsection{Phase difference}
\label{sec:res_pha}

From \eqref{eq:phase} one may analyse the phase difference of the oscillations between the different time series. 
For the frequency region where we see the solar-like oscillations as trapped standing waves the adiabatic expectation is a phase difference of $\Delta\phi=0^{\circ}$.
Non-adiabatic effects (\ie, damping and excitation) will cause non-zero phase differences. Typically such analyses have, however, been done between velocity and photometry observations \citep[see, \eg,][]{1999A&A...341..625P,1999ApJ...525.1042J,2004ApJ...616..594S,2006ESASP.624E.101S,2008AN....329..498S}.

\begin{figure}
\includegraphics[width=\columnwidth]{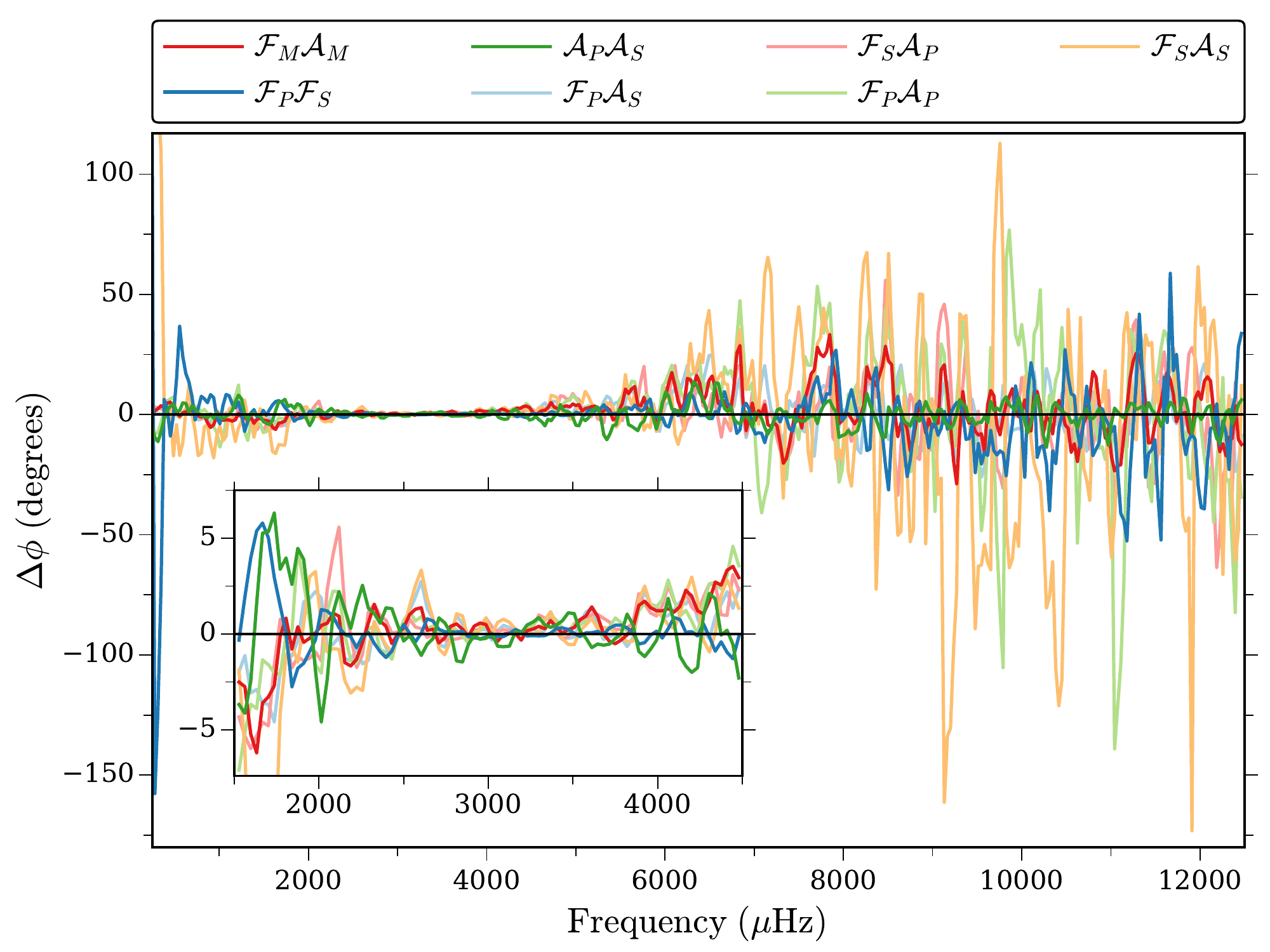}
\caption{Phase difference $\Delta\phi$ in degrees between the different 2B time series combinations as a function of frequency. The legend indicates first the cell, forward ($\mathcal{F}$) or aft ($\mathcal{A}$), and then by the subscript the detector, \ie, the port ($P$), starboard ($S$), or mean ($M$) detector data products, used in the different time series combinations. The insert shows a zoom-in of the frequency region dominated by oscillations.}
\label{fig:phase}
\end{figure}

The phase differences for the different cell and detector combinations are shown in \fref{fig:phase}. We have omitted adding uncertainties to the phase functions, but these may be computed according to the method outlined by \citet[][]{1999ApJ...525.1042J}. As expected, the phase differences in the region dominated by the stellar oscillations (${\sim}1500$ to ${\sim}4500\, \rm \mu Hz$) are indeed close to $0^{\circ}$.
Some minor differences are seen between the different phase functions, specifically it is seen that the $\mathcal{F}_P\mathcal{F}_S$ and $\mathcal{A}_P\mathcal{A}_S$ combinations are better centred on $\Delta\phi=0^{\circ}$ compared to, for instance, the $\mathcal{F}_M\mathcal{A}_M$ combination. Because the two former combinations probe the atmosphere at two spatially differently weighted regions, but, crucially, at the same atmospheric height, they should be expected to give $\Delta\phi=0^{\circ}$. The latter, on the other hand, isolates the phase difference from different heights, and is seen to be well followed by the other combinations that also probe different heights. The slight upward trend in the phase difference, approaching $\Delta\phi\approx3^{\circ}$ near the acoustic cut-off at $\rm {\sim}5000\, \mu Hz$, could be the signature of non-adiabatic effects. A contribution could also come from high frequency travelling waves above the trapping region that do not maintain their phase with height, as do standing waves. The increase of phase-shift with height will for this contribution depend on frequency.

As in a Fourier transform, phase leakage in the transform (\eqref{eqn:summer}) can affect the calculated phase difference (\eqref{eq:phase}), especially at frequencies of low S/N.
It is also important to remember that the frequency resolution per design is low, such that, within a given frequency bin the average phase from all modes of oscillations in the bin is obtained (with different angular degree $l$). 
In addition, signals from the granulation will, as in the power density spectrum, have a component in the frequency region dominated by the oscillations. In constrast to the spectral power density, it is difficult to isolate this component in the phase signal. However, assuming a random phase of this signal, due to the stochastic nature of granulation, between $\pm 180^{\circ}$, the larger number of measurements per bin and regression towards the mean would suggest a contribution that is close to $\Delta\phi=0^{\circ}$. 

In the granulation-dominated frequency region (${\lesssim}1500\, \rm \mu Hz$), and at high frequencies (${\gtrsim}4500\, \rm \mu Hz$), where both pseudo-modes \citep[][]{2005ApJ...623.1215J} and white noise contribute to the signal, the phase differences are centred around $0^{\circ}$, but with larger deviations. This suggests a randomised phase, as expected from incoherent signals.

\subsection{Cross spectrum}
\label{sec:res_cro}

\begin{figure*}
\includegraphics[width=\textwidth]{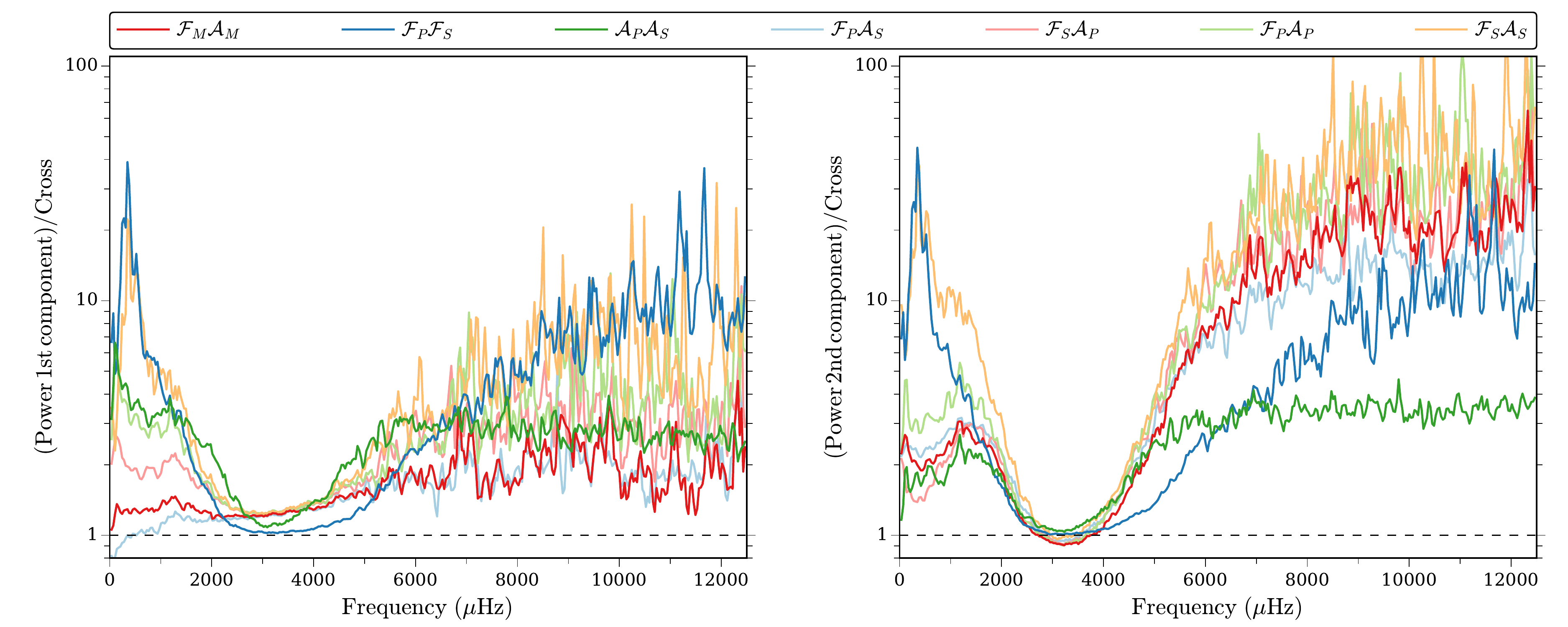}
\caption{Comparison between the power and cross spectra for the different 2B time series combinations as a function of frequency. The legend indicates first the cell, forward ($\mathcal{F}$) or aft ($\mathcal{A}$), and then by the subscript the detector, \ie, the port ($P$), starboard ($S$), or mean ($M$) detector data products, used in the different combinations. Left: ratio between power spectrum from the first component of the specific combination of time series used for the cross spectrum. As an example, the dark red line thus gives the comparison between the power spectrum from $\mathcal{F}_M$ (the 1st component in $\mathcal{F}_M\mathcal{A}_M$) and the cross spectrum from $\mathcal{F}_M\mathcal{A}_M$. Right: ratio between power spectrum from the second component of the specific combination of time series used for the cross spectrum. As an example, the dark red here then gives the comparison between the power spectrum from $\mathcal{A}_M$ (the 2nd component in $\mathcal{F}_M\mathcal{A}_M$) and the cross spectrum from $\mathcal{F}_M\mathcal{A}_M$. Except for the $\mathcal{F}_P\mathcal{F}_S$ and $\mathcal{A}_P\mathcal{A}_S$ combinations, the left (right) panel shows the comparisons between power spectra from the forward (aft) cell data with the cross spectrum.}
\label{fig:cross}
\end{figure*}

It is in the cross spectrum that the effect from the incoherency of granulation noise must be found for the improved detection of oscillations.
\fref{fig:cross} shows the comparisons between power and cross spectra from different data product combinations from the forward and/or aft cells.

As seen, an improvement is indeed gained from using the cross spectrum
compared to the power spectra from any individual data product.  Of
particular interest is the $\mathcal{F}_M\mathcal{A}_M$ combination,
which in the cross spectrum best isolates the effect of combining
different heights. The typical BiSON product is given by the power
spectrum from $\mathcal{F}_M$. For many of the combinations we see an
improvement that exceeds that possible by merely averaging the data in
the temporal domain, where a maximum improvement by a factor of two
(in power) is possible by combining fully incoherent Gaussian
noise. We note that even though the incoherent noise is reduced, so is
the frequency resolution when using Welch's method --- this will
impact the detection of these low frequency modes that have long
lifetimes, hence narrow line widths. In a future paper we will
investigate the possibility of a different reduction of BiSON data
from a single cell, which currently adopts a temporal averaging of
data from the port and starboard detectors, and the potential gain in
detection of low frequency modes from the cross spectrum of 2B data
using Welch's and the smoothing methods \citep[][]{Pardo2012207}.
From \fref{fig:cross} it is furthermore evident that the quality of
data from the forward cell exceeds that from the aft cell --- from the
$\mathcal{F}_M\mathcal{A}_M$ data we see, for instance, a
significantly larger improvement from using the cross spectrum from
both cells versus the aft cell power spectrum compared to that from
the forward cell.

\section{Discussion}
\label{sec:dis}

We have demonstrated that the granulation signal from different atmospheric heights in the solar atmosphere is not coherent, and shown the potential for reducing the low frequency 
granulation-noise in the cross-spectrum, increasing the signal-to-noise of the coherent modes. This will enable the positive detection of additional low frequency modes.
The results here will be used in a future study of data from the radial velocity instrument of the Stellar Observations Network Group (SONG) to improve 
observations of both the Sun and solar-like oscillators in general.
Our study also showed a phase difference from different heights as expected from theory, \ie, near the asymptotic value of a $0^{\circ}$ difference for the region dominated by oscillations, and randomised in the low- and high-frequency noise-dominated regions.  

The results covered here also support a new network concept called
BiSON-Mini. The idea is to deploy many compact and affordable
versions of the BiSON instrument around the globe. The high number of
nodes will ensure monitoring of the Sun as-a-star with a coverage of
close to $100\%$, effectively reducing noise and artefacts in the
power spectrum from temporal gaps; BiSON generally achieves a coverage
just over $80\%$. The BiSON-Mini nodes would, from their different
longitudinal positions and hence slightly different line-of-sight rotational velocities with respect to the Sun, probe different depths in the solar
photosphere.  Combining contemporaneous data from many such nodes
would thus benefit strongly from the observed incoherence of the
granulation noise, as shown in this study, in addition to minimising
other incoherent noise contributions, such as that from the earth
atmosphere or instrumental signals. This would enable the detection of
additional low frequency solar oscillation modes.

In a future paper we will use the 2B data to study the depth dependence of seismic and granulation properties in the Sun, which will serve as the first step to a similar study using SONG data.

\section*{Acknowledgments}
\footnotesize
We thank the referee for many useful comments and suggestions that helped to improve the paper.
We would like to thank all those who are, or have been, associated with BiSON. BiSON is funded by the Science and Technology Facilities Council (STFC).
MNL acknowledges the support of The Danish Council for Independent Research | Natural Science (Grant DFF-4181-00415).
WJC, SJH, GRD, YPE, and RH acknowledge the support of the UK Science and Technology Facilities Council (STFC). 
Funding for the Stellar Astrophysics Centre (SAC) is provided by The Danish National Research Foundation (Grant DNRF106). 

\small

\bibliography{MasterBIB}
\label{lastpage}

\end{document}